\documentclass{INTERSPEECH2023}
\usepackage[table]{xcolor}
\usepackage{adjustbox}
\usepackage{multirow}
\usepackage[symbol]{footmisc}
\renewcommand{\thefootnote}{\fnsymbol{footnote}}
\newcommand\blfootnote[1]{%
  \begingroup
  \renewcommand\thefootnote{}\footnote{#1}%
  \addtocounter{footnote}{-1}%
  \endgroup
}
\newcolumntype{C}[1]{>{\centering}m{#1}}
\interspeechcameraready 

\title{Brouhaha: multi-task training for voice activity detection, \\ speech-to-noise ratio, and C50 room acoustics estimation}
\name{Marvin Lavechin$^{\star,1,2}$, Marianne Métais$^{\star,1}$, Hadrien Titeux$^1$, Alodie Boissonnet$^2$, Jade Copet$^2$, Morgane Rivière$^2$, Elika Bergelson$^3$, Alejandrina Cristia$^1$, Emmanuel Dupoux$^{1,2}$, Hervé Bredin$^4$}

\address{
$^1$ LSCP, DEC, ENS, EHESS, CNRS, PSL University, Paris, France 
\\
$^2$ Meta AI Research, France
\quad
$^3$ Duke University, North Carolina, USA 
\\
$^4$ IRIT, Université de Toulouse, CNRS, Toulouse, France}

\email{marvinlavechin@gmail.com}

\begin{document}
\maketitle
\blfootnote{$^{\star}$ M. Lavechin and M. Métais equally contributed to this work.}
\blfootnote{This work was granted access to the HPC resources of GENCI-IDRIS under the allocation 2022-AD011012554. It also benefited from the support of ANR-16-DATA-0004 ACLEW,  ANR-17-EURE-0017, ANR-19-P3IA-0001; the J. S. Mc-Donnell Foundation; and ERC ExELang grant no 101001095.}

\begin{abstract}
Most automatic speech processing systems register degraded performance when applied to noisy or reverberant speech. But how can one tell whether speech is noisy or reverberant? We propose Brouhaha, a neural network jointly trained to extract speech/non-speech segments, speech-to-noise ratios, and C50 room acoustics from single channel recordings. Brouhaha is trained using a data-driven approach in which noisy and reverberant audio segments are synthetized. We first evaluate its performance and demonstrate that the proposed multi-task regime is beneficial. We then present two scenarios illustrating how Brouhaha can be used on naturally noisy and reverberant data: 1) to investigate the errors made by a speaker diarization model (pyannote.audio); and 2) to assess the reliability of an automatic speech recognition model (Whisper from OpenAI). Both our pipeline and a pretrained model are open source and shared with the speech community.
\end{abstract}


\section{Introduction and related work}

\label{sec:intro}

Robustness to degraded acoustic environments is a critical factor limiting the impact and adoption of speech technologies. Numerous sources of variations in the audio can degrade or hide the signal of interest and impact the performance of automatic speech processing systems. Be it automatic speech recognition (ASR) \cite{giri2015improving,kinoshita2016summary,gamper2020predicting}, speaker identification/diarization \cite{zhao2014robust,ryant2020third}, or speaker localization \cite{chakrabarty2017broadband}, most systems exhibit a loss of performance when applied in noisy or reverberant conditions.

While speech processing systems are being improved to handle degraded acoustic environments \cite{ko2017study,donahue2018exploring,narayanan2023learning}, little work has been devoted to automatically predict the properties of the acoustic environment. A proposed approach involves using synthetic audio generated by applying an audio transformation of interest (e.g., reverberation). A neural network is then trained to extract the `strength` of this audio transformation. This approach is most commonly used to develop systems that predict room acoustic measures like speech clarity ($C_{50}$), reverberation time ($T_{60}$) or direct-to-reverberant ratio (DRR) \cite{eaton2015ace,parada2016single,xiong2018exploring,bryan2020impulse,gamper2020blind}. In practice, these values can be estimated directly from the room impulse response (RIR, the recording of a high-energy and bursty sound, such as a pistol shot or a balloon popping). However, in most cases, RIRs are not available, and we need to estimate the values of interest from the observed single channel audio recording. A similar approach has been adopted in \cite{li2020frame} to automatically estimate the frame-level speech-to-noise ratio (SNR). The authors evaluate the performance of their system on synthetic data, but not on real data. In practice, real SNRs are not available making it impossible to compare the estimated values to the real ones. Thus, it remains unclear if such a system can generalize to real data.

Given the high interplay between noise and reverberation (the SNR may be influenced by how noise and speech sources reverberate, and it is harder to obtain reliable estimates of reverberation parameters in low SNR conditions \cite{lollmann2019comparative,eaton2016estimation}), can we design a system that tackles both tasks simultaneously? This is one of the questions we address in this work. Our approach is closest to \cite{looney2020joint} who proposes to train a neural network for jointly estimating room acoustic parameters and the utterance-level SNR. However, the authors use a restrained set of noise segments which cast doubts on the ability of their model to generalize to unseen noises. More importantly, they do not evaluate their system with respect to the SNR, and they do not address the question of whether the proposed multi-task regime is beneficial for the estimation performance.

We propose \textit{Brouhaha}, a model jointly trained on the speech/non-speech classification task and the SNR and $C_{50}$ regression tasks. Our model is trained on $1,250$ hours of synthetic audio generated from clean speech segments contaminated with silence, noise and reverberation. We first demonstrate that the proposed multi-task regime is beneficial and compare the performance of \textit{Brouhaha} against state-of-the-art systems. We then apply \textit{Brouhaha} on real data (under naturally noisy and reverberant conditions) to: 1) analyze the error patterns of a speaker diarization system (\textit{pyannote.audio} \cite{bredin2020pyannote}); and 2) assess the reliability of an ASR system (Whisper \cite{radfordrobust}). In addition to showing how \textit{Brouhaha} can be used, these experiments constitute evidence that our system is applicable to real data.

Beyond the scientific interest of exploring the effectiveness of the proposed multi-task training regime and assessing the applicability of the method on real data after training on synthetic ones, we believe our work has a strong practical interest. Unlike previous work \cite{li2020frame,looney2020joint}, \textit{Brouhaha} can be applied to any audio regardless of whether it contains speech, non-speech or both. By using our system, there is no requirement to implement a preliminary voice activity detection system prior to obtaining SNR and C50 values. We believe such advancement, in addition to a simple user interface (one python command!), significantly aids empowering researchers who may not possess expertise in speech processing or machine learning to make the most out of speech technology.

\begin{figure*}[htb]
    \centering
    \includegraphics[width=.85\textwidth, trim={0.9cm 16.7cm 2.8cm 0.2cm}, clip]{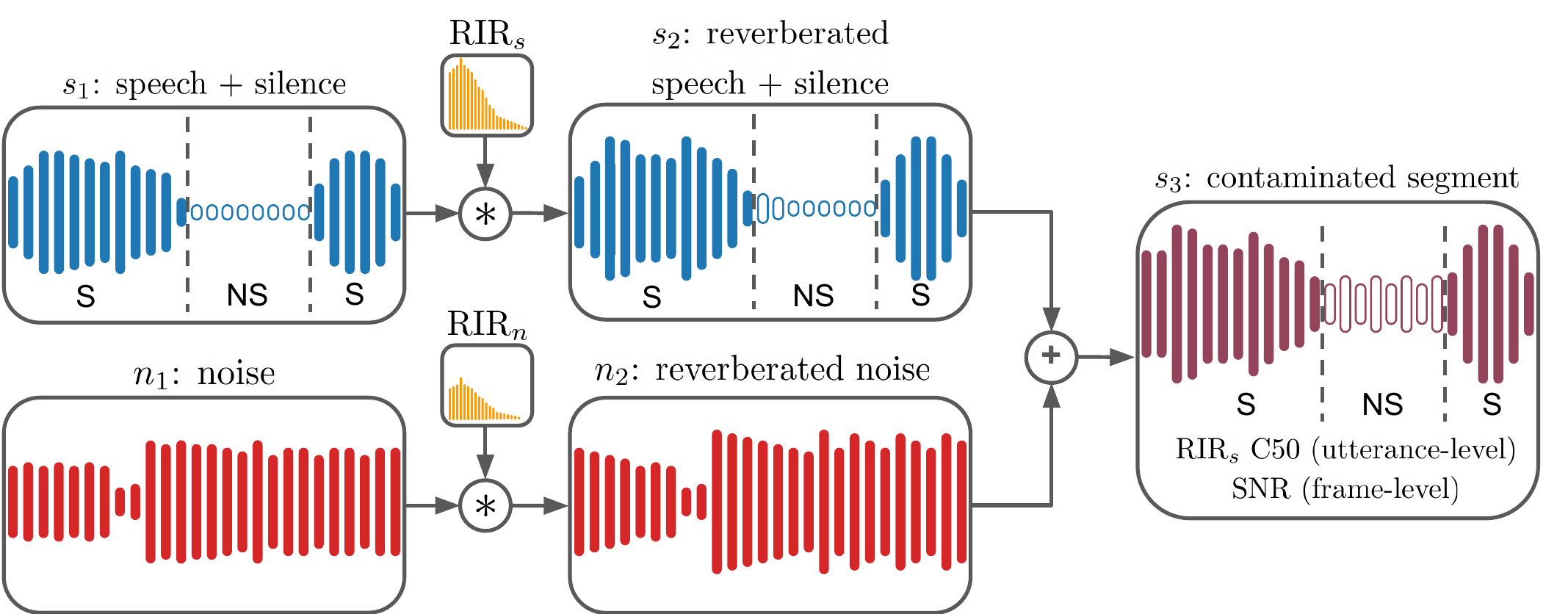}
    \caption{\textbf{Audio contamination pipeline.} $s_1 \rightarrow s_2$: With probability $p_{\text{RIR}}=0.9$, the clean speech segment (marked as S) contaminated with silence (marked as NS) $s_1$ is convolved with a randomly drawn impulse response $\text{RIR}_s$. $n_1 \rightarrow n_2$: With probability $p_{\text{RIR}}$, the randomly drawn noise segment $n_1$ is convolved with a randomly drawn impulse response $\text{RIR}_n$. $s_2 + n_2 \rightarrow s_3$: The reverberated speech segment $s_2$ and the reverberated noise segment $n_2$ are added together to obtain a Speech-to-Noise Ratio (SNR) randomly drawn between $0$ and \SI{30}{\decibel}. As noises can have a wide dynamic range and the utterance-level SNR captures only global information about the noise level, we recompute SNRs using a 2-second long sliding window shifted every \SI{10}{\milli\second} over $s_2$ and $n_2$. $C_{50}$ is computed as the ratio of early ($0$ to \SI{50}{\milli\second}) and late (\SI{50}{\milli\second} to the end of the response) energies of the room impulse response $RIR_s$. Labels obtained via this pipeline include: speech/non-speech (frame-level), $C_{50}$ measure of $\text{RIR}_s$ (utterance-level), and SNR (frame-level).
    }
    \label{fig:pipeline}
\end{figure*}

\section{Audio contamination pipeline}

We start from: 1) a set of clean speech segments that will be contaminated; 2) a set of noise segments used to simulate noisy conditions; and 3) a set of RIRs to simulate reverberation. The clean speech segments are contaminated following the steps presented in Figure \ref{fig:pipeline}, which we will not repeat here.  

\section{Multi-task training}

We tackled the voice activity detection problem as a classification problem where, for each $16$-{\unit{\milli\second}} frame, the expected output is 1 if there is speech, 0 otherwise. $C_{50}$ and SNR estimations were tackled as regression problems where, for each $16$-{\unit{\milli\second}} frame, the expected output is the actual $C_{50}$ or SNR in {\unit\decibel}. We tackled the $C_{50}$ estimation at the frame level during training -- despite the label being at the utterance level -- to allow the model to return smoother transitions when a change in $C_{50}$ is detected at inference time.

At training time, short fixed length sub-sequences are drawn randomly from the training set and gradient-descent is used to minimize the multi-task loss function $\mathcal{L} = \mathcal{L}_{\text{VAD}} + \mathcal{L}_{C_{50}} +  \mathcal{L}_{\text{SNR}}$, where $\mathcal{L}_{\text{VAD}}$ is the binary cross-entropy loss, and $\mathcal{L}_{C_{50}}$ and $\mathcal{L}_{\text{SNR}}$ are mean squared error (MSE) losses. Before training, $\mathcal{L}_{C_{50}}$ and $\mathcal{L}_{\text{SNR}}$ are normalized by their maximum value (computed over 10 batches) to ensure all three losses lie between 0 and 1. We computed $\mathcal{L}_{\text{SNR}}$ only over speech frames as the SNR is not defined on non-speech frames.

\section{Experiments}
\label{sec:exp}
\subsection{Datasets}

Our audio contamination pipeline requires three types of audio data: 1) clean speech segments; 2) noise segments; and 3) RIRs. A pretrained VAD model~\cite{bredin2020pyannote} was applied to find non-speech segments in $1000$ hours of clean read-speech, retrieved from the LibriSpeech~\cite{panayotov2015librispeech}. Predicted non-speech segments were extended with silence to obtain a ratio of approximately \SI{30}{\percent} of non-speech. We used noise segments from AudioSet~\cite{gemmeke2017audio} and discarded human vocalizations. We also downsampled music segments from \SI{38}{\percent} to \SI{5}{\percent}, leading to a total of $1500$ hours of noise segments. Finally, $385$ impulse responses were obtained from EchoThief \cite{warrenechothief} and the MIT Acoustical Reverberation Scene \cite{traer2016statistics} datasets. We used the same train/dev/test split originally proposed in LibriSpeech. Noise segments and impulse responses were randomly split into \SI{80}{\percent}, \SI{10}{\percent} and \SI{10}{\percent} for the training, development and test set, respectively. All files used in this paper consist of 16-kHz single-channel recordings.

\subsection{Evaluation metrics}
\label{sec:metrics}

We evaluated \textit{Brouhaha} performance on the VAD task using the F-score between precision and recall, such as implemented in \textit{pyannote.metrics} \cite{pyannote.metrics}. SNR and $C_{50}$ predictions were evaluated using the mean absolute error (MAE) at the frame level. Since SNR is not defined on non-speech frames, the SNR was only evaluated across speech frames.

\subsection{Architecture, optimization and training procedure}
\label{sec:hyperparam}

The model consists of SincNet (using the configuration in \cite{sincnet}), followed by a stack of bidirectional long short-term memory (LSTM) and feed-forward layers. Finally, we have three parallel layers: one classification layer (with \textit{softmax} activation) that returns the predicted probability of speech, and two regression layers that return the predicted SNR and $C_{50}$ (with \textit{sigmoid} activation parametrized between \SI{-15} and \SI{80}{\decibel} for the SNR, and \SI{-10} and \SI{60}{\decibel} for the $C_{50}$).

We trained $144$ different architectures across different sets of hyperparameters, varying the duration of the input sequences: $4$, $6$, $8$, or $10$ seconds; the batch size: $32$, $64$, or $128$ sequences; the size of the hidden LSTM layers: $128$ or $256$; the number of LSTM layers: $2$ or $3$; and the dropout proportion: $0$, $30$ or \SI{50}{\unit{\percent}}. The best architecture was trained with $6$-{\unit\second} segments, a batch size of $64$ sequences, $3$ LSTM layers of size $256$, and a dropout proportion of \SI{50}{\unit{\percent}}. The best architecture was selected on the validation metric: an average of the VAD F-score, SNR and $C_{50}$ MAEs, with the latter two normalized by the maximum error to balance the contribution of each term. 

\section{Results}
\label{sec:results}

\subsection{The effect of multi-task training}

\begin{table}[tbh]
  \centering
  \caption{Performance on unseen synthetic data (our test set) in terms of F-score (VAD) and mean absolute errors (SNR and $C_{50}$). A checkmark below a given training task indicates that the associated loss is activated during training.}
  \begin{tabular}{C{.5cm} C{.5cm} C{.5cm} C{1.3cm} C{1.3cm} c}
    \toprule
    \multicolumn{3}{c}{\textbf{Training tasks:}} & \textbf{VAD} & \textbf{SNR} & $\mathbf{C_{50}}$ \\
    \textbf{VAD} & \textbf{SNR} & $\mathbf{C_{50}}$ &
    \shortstack{\textbf{F-score ({\boldmath \unit{\percent}})}} & 
    \shortstack{\textbf{MAE ({\boldmath \unit{\decibel}})}} & 
    \shortstack{\textbf{MAE ({\boldmath \unit{\decibel}})}}
    \\
    \midrule
    \checkmark & \checkmark & \checkmark & $\mathbf{93.7}$ & $\mathbf{4.1}$ & $\mathbf{3.5}$ \\
    \checkmark & \checkmark &  & $\mathbf{93.7}$ & $4.2$ & ------\\
    \checkmark & & \checkmark & $93.6$ & ------ & $3.8$ \\
    & \checkmark & \checkmark & ------ & $4.3$ & $3.7$ \\
    \checkmark & & & $93.5$ & ------ & ------ \\
    & \checkmark & & ------ & $4.3$ & ------ \\
    & & \checkmark & ------ & ------ & $4.2$ \\
    \bottomrule
  \end{tabular}
    \label{tab:multitask}
\end{table}

\noindent Table \ref{tab:multitask} shows performance obtained by models trained to solve either one, two or three of the proposed tasks (VAD, SNR, $C_{50}$). All models shared the same set of hyper-parameters, only the dimension of the output layer differed. Results indicate that the multi-task training regime is beneficial: the model trained simultaneously on the three tasks obtained better performance than models trained on two tasks which themselves obtained better performance than models trained on a single task. The largest performance gain is observed for the $C_{50}$ estimation, with a decrease of \SI{0.7}{\decibel} in terms of MAE between the single-task and the three-tasks training regime. These results seem to show that sharing weights during training helps better solve the proposed three tasks. Not only does using a single model provide a performance gain, but it is also more convenient and computationally efficient.

\subsection{Voice activity detection}
\label{sec:vad}

\begin{table}[tbh]
  \centering
  \caption{Voice activity detection F-score obtained by \textit{Brouhaha} and \textit{pyannote.audio} pretrained system \cite{bredin2020pyannote}. Numbers are reported on synthetic data (our test set) and on real data (BabyTrain \cite{Lavechin2020AnOV}).} 
  \begin{tabular}{llc}
    \toprule
    \textbf{Data type} & \textbf{System} & \textbf{VAD F-score ({\boldmath \unit{\percent}})}
    \\
    \midrule
    \multirow{2}{*}{synthetic} & Brouhaha (ours) & $\mathbf{93.7}$ \\
    & pyannote.audio \cite{bredin2020pyannote} & $89.0$ \\
    \midrule
    \multirow{2}{*}{real} & Brouhaha (ours) & $77.2$ \\
    & pyannote.audio \cite{bredin2020pyannote} & $\mathbf{80.8}$ \\
    \bottomrule
  \end{tabular}
    \label{tab:vad}
\end{table}

\noindent Table \ref{tab:vad} shows voice activity detection performance obtained by \textit{Brouhaha} and a state-of-the-art system (\textit{pyannote.audio} \cite{bredin2020pyannote}). We consider two evaluation sets: 1) our test set made of unseen synthetic audio data (referred as `synthetic` in the table); and 2) BabyTrain \cite{Lavechin2020AnOV}, a corpus of highly naturalistic child-centered recordings (referred as `real` in the table). Specifically, BabyTrain recordings are acquired via child-worn microphones as they go about their everyday activities and are widely used in language acquisition research \cite{lavechin2022reverse}. Child-centered recordings are notoriously challenging for speech processing systems as they contain spontaneous and overlapping speech, and a wide variety of noisy and reverberant conditions.

Results show a strong advantage for \textit{Brouhaha} over \textit{pyannote.audio} on unseen synthetic data (\SI{4.7}{\percent} absolute difference in terms of F-score) . This indicates that, on highly noisy and reverberant synthetic audio, our system is competitive on the VAD task. Admittedly, \textit{Brouhaha} has an advantage over \textit{pyannote.audio} as the latter has not been trained on synthetically noisy and reverberant audio. Turning to a performance comparison on real data, numbers reveal that \textit{pyannote.audio} outperforms \textit{Brouhaha} by a \SI{3.6}{\percent} absolute difference in terms of F-score. This result suggests that training a VAD system on LibriSpeech  \cite{panayotov2015librispeech} contaminated with reverberation and additive noise might not be optimal, and this is despite the precautions taken in simulating challenging noisy and reverberant conditions. Nonetheless, LibriSpeech is currently the only source of clean speech available in sufficiently large quantities to run our audio contamination pipeline and obtain SNR and $C_{50}$ labels.

\subsection{Speech-to-noise ratio estimation}

\begin{table}[tbh]
  \centering
  \caption{Mean absolute error on the SNR estimation task computed on unseen synthetic data (our test set). All predicted and gold SNRs are brought back to the $[-15,30]$ {\unit\decibel} range as done in \cite{li2020frame}. For a given speech utterance, the heuristic estimates the noise (resp. speech) power as the mean power of non-speech (resp. speech) frames within a $6$-{\unit\second} window centered around each annotated speech frame (defaulting to the average SNR when no non-speech frames were found within the $6$-{\unit\second} window).} 
  \begin{tabular}{lc}
    \toprule
    \textbf{System} & \textbf{SNR MAE ({\boldmath \unit{\decibel}})}
    \\
    \midrule
    Brouhaha (ours) & $\mathbf{2.3}$ \\
    Heuristic & $8.4$ \\
    Li et al. \cite{li2020frame} & $12.5$ \\
    \bottomrule
  \end{tabular}
  \label{tab:snr}
\end{table}

\noindent Table \ref{tab:snr} shows MAE performance on the SNR estimation task computed on our test set made of unseen synthetic audio data for: 1) \textit{Brouhaha}; 2) a heuristic using the oracle VAD that estimates the noise (resp. speech) as the mean power of neighboring non-speech (resp. speech) frames; and 3) the system proposed in \cite{li2020frame} (a $4$-layer LSTM trained from mel frequency cepstral coefficients). 

Results indicate that \textit{Brouhaha} is better at estimating the frame-level SNR than our heuristic, with an absolute difference of \SI{6.1}{\decibel} in terms of MAE (note that both systems use a $6$-{\unit\second} window as input, and that our heuristic requires oracle VAD boundaries). Surprisingly, our heuristic performs better than the system proposed in \cite{li2020frame} with a \SI{4.1}{\decibel} absolute difference in terms of MAE. This indicates that \cite{li2020frame} struggles generalizing to unseen noise or to reverberant environments. Unfortunately, we could not compare systems on the test used in \cite{li2020frame} as the latter has not been publicly released.

\subsection{$C_{50}$ estimation}

\begin{figure}[tbh]
\centering
\includegraphics[width=.75\columnwidth]{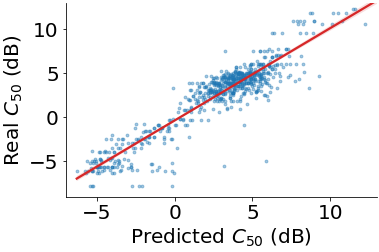}
\caption{\textbf{$\mathbf{C_{50}}$ estimation.} Real $C_{50}$ against $C_{50}$ predicted by \textit{Brouhaha} on 1000 utterances from the BUT Speech@FIT Reverb dataset~\cite{szoke2019building}.}
\label{fig:c50}
\vspace{-0.5cm}
\end{figure}

\noindent We ran \textit{Brouhaha} on the BUT Speech@FIT Reverb dataset~\cite{szoke2019building}. This dataset consists of LibriSpeech test-clean utterances retransmitted by a loudspeaker in $5$ different rooms. For each room, the speaker was placed on $5$ positions on average and retransmitted utterances were recorded with $31$ microphones. RIRs were measured multiple times for each speaker position. Here, we compare the real~$C_{50}$ (averaged over between 1 and 9 duplicated RIR measures) to the $C_{50}$ predicted by \textit{Brouhaha} on $1000$ randomly drawn utterances.

Figure \ref{fig:c50} shows a strong correlation between the real and the predicted~$C_{50}$, with a $R^2$ of $.85$ and a mean average error of \SI{1.1}{\decibel}. We would have liked to compare the performance of our system on the $C_{50}$ estimation task with other systems, but we could not find any open-source pre-trained $C_{50}$ estimators despite extensive research in this area \cite{parada2016single,xiong2018exploring,gamper2020blind}.


\subsection{Investigating speaker diarization errors}

\begin{figure}[tbh]
\centering
\includegraphics[width=\columnwidth,trim={0 0.3cm 0 0.55cm}, clip]{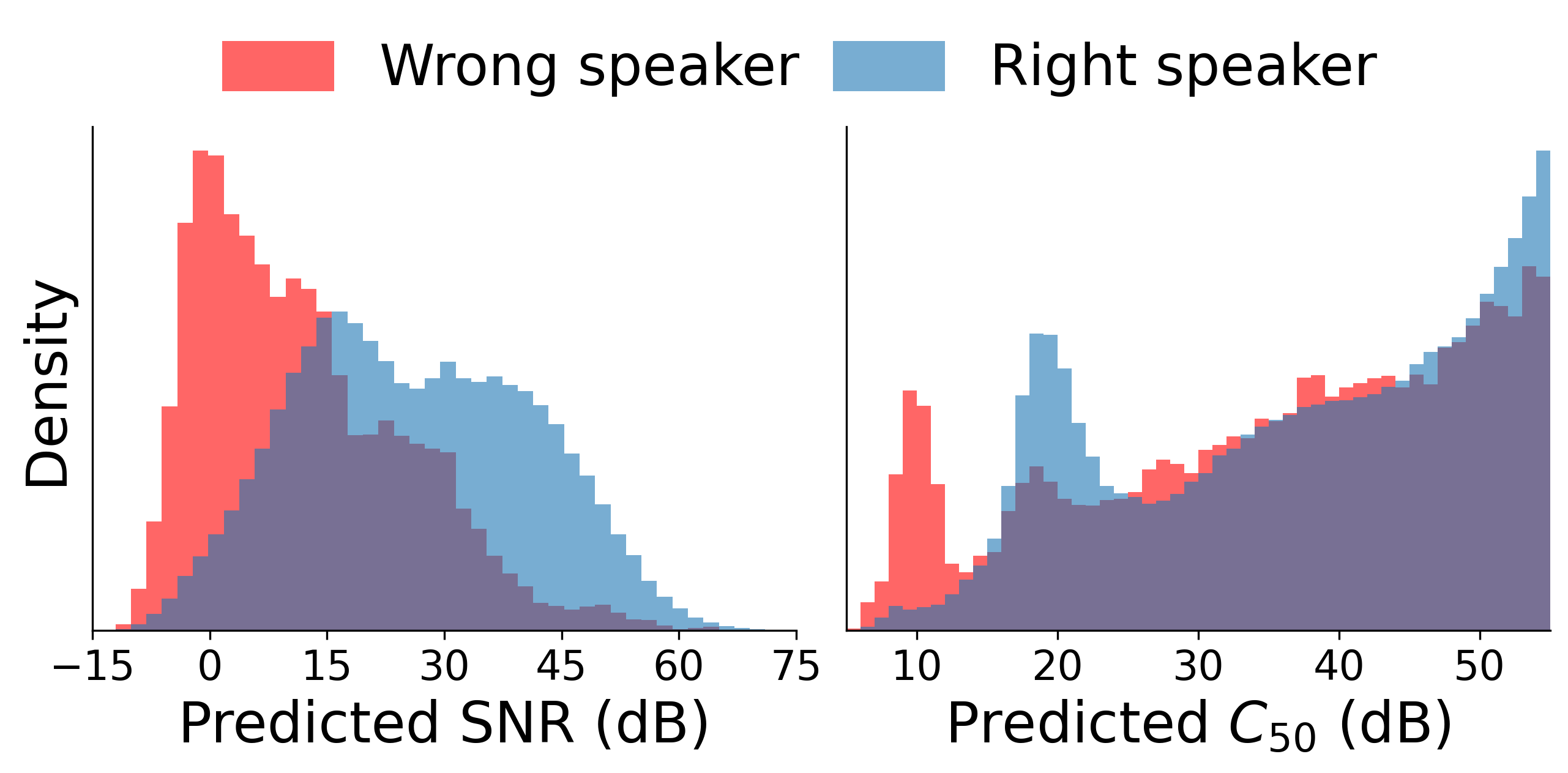}
\caption{\textbf{Investigating speaker diarization errors.} Distribution of SNR (left) and $C_{50}$ (right) predicted by \textit{Brouhaha} as a function of whether a pretrained speaker diarization system \cite{bredin2020pyannote} assigns a speech frame to a wrong (red) or to the right speaker (blue).}
\label{fig:diarization}
\vspace{-0.5cm}
\end{figure}

\noindent We ran a pretrained \textit{pyannote.audio} speaker diarization pipeline~\cite{bredin2020pyannote} on the VoxConverse dataset~\cite{VoxConverse} and evaluated its performance at \textit{Brouhaha} frame resolution (\SI{16}{\milli\second}). Each frame can either be classified as: 1) missed detection (when the speaker diarization pipeline incorrectly classifies a speech frame as non-speech): 2) false alarm (the other way around); 3) speaker confusion (when a speech frame is assigned to the wrong speaker); or 4) correct. Figure~\ref{fig:diarization} focuses on speaker confusion (but the same pattern holds for missed detections) and shows the distribution of predicted SNR (left) and $C_{50}$ (right) depending on whether the speech frame was assigned to the correct speaker. There is a clear trend as far as SNR is concerned: \textit{pyannote.audio} is much more likely to confuse speakers in low (predicted) SNR regions. Similarly, the accuracy degrades significantly as we get closer to the lowest predicted $C_{50}$ values.

Exploring the errors made by a pretrained system can provide valuable insights for developing effective strategies. In our case, one might devise strategies to address the issue of high speaker confusion in low SNR conditions: increasing the weight of low-SNR sequences in the training loss, or running speech enhancement algorithms on low SNR areas for instance.

\subsection{Assessing the reliability of an ASR system}

We ran Whisper large ASR system \cite{radfordrobust} on highly naturalistic speech utterances from the American English Bergelson corpus \cite{bergelson2019north,seedlingsdata} (child-centered recordings, similar to the ones used in Section \ref{sec:vad}). We evaluate the performance of Whisper using the percentage hits (i.e., percentage of words correctly transcribed). We include a total of 804 utterances at least $5$-words long (as short sequences most often led to a score \SI{0}{\percent} or \SI{100}{\percent}).

\begin{figure}[tbh]
\centering
\includegraphics[width=.85\columnwidth,trim={0 0.29cm 0 0.52cm}, clip]{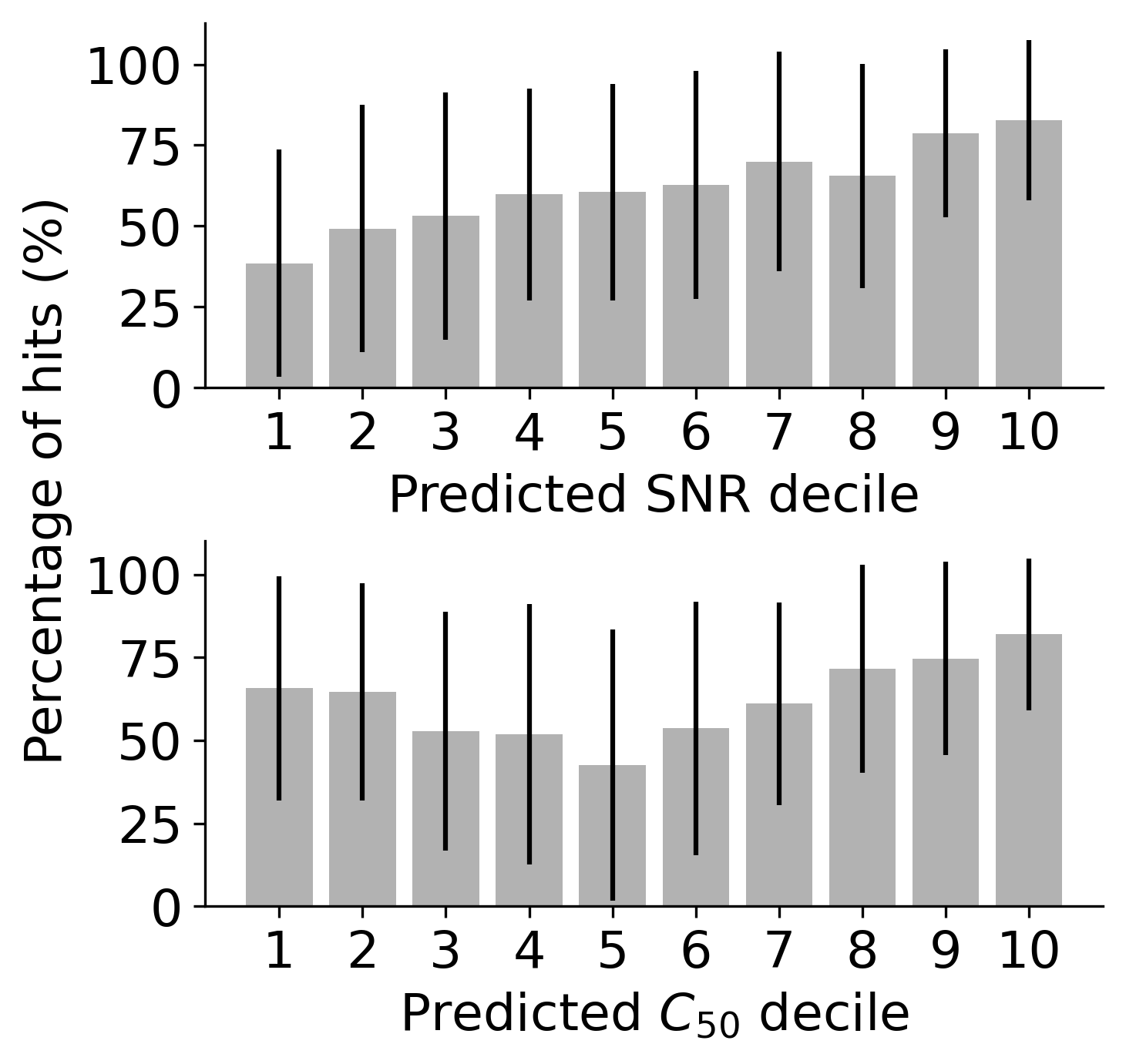}
\caption{\textbf{Assessing the reliability of an ASR system.} Percentage of hits obtained by Whisper large as a function of predicted SNR decile (top panel) and predicted $C_{50}$ decile (bottom panel). Bars represent the percentage of hits averaged across utterances. Thin black lines represent standard errors.}
\label{fig:seedlings}
\end{figure}

Figure \ref{fig:seedlings} shows the average percentage of hits obtained by Whisper for utterances binned according to their predicted SNR (top panel) or $C_{50}$ (bottom panel) decile. On average, Whisper correctly transcribes \SI{83}{\percent} of the words on utterances whose SNR belongs in the $[12,24]$~{\unit\decibel} (last SNR decile, top panel). This number decreases as the SNR decreases until Whisper successfully transcribes only \SI{38}{\percent} of the words on utterances whose SNR is in the $[-9,-4]$~{\unit\decibel} range (first SNR decile). Although utterances whose predicted $C_{50}$ is high tend to be better transcribed by Whisper, the trend with respect to the $C_{50}$ is less clear (bottom panel). In conclusion, by using \textit{Brouhaha}, we demonstrated the low reliability of Whisper on noisy utterances found in child-centered long-forms.

\section{Conclusion and future work}

We proposed \textit{Brouhaha}, a model jointly trained on the voice activity detection, SNR, and $C_{50}$ estimation tasks. After evaluating the performance of our system and demonstrating that the multi-task training regime is beneficial, we illustrated two use cases showing how our model can be used on real data. Beyond investigating errors made by speech processing systems or assessing their reliability in noisy and reverberant conditions, we foresee other potential downstream tasks, e.g., SNR- or $C_{50}$-based microphone selection \cite{wolf2009towards} or SNR-aware speech enhancement \cite{fu2016snr}. Future work could explore these downstream tasks, the use of spontaneous clean speech to improve VAD performance, or the estimation of other room acoustic parameters, such as $T_{60}$ or DRR. Both a pre-trained model and our audio contamination pipeline are shared with the community\setcounter{footnote}{0}\footnote{\texttt{https://github.com/marianne-m/brouhaha-vad}}.

\vfill\pagebreak

\ninept

\bibliographystyle{IEEEtran}
\bibliography{mybib}

\end{document}